\documentclass[conference, 10pt]{IEEEtran}
\IEEEoverridecommandlockouts

\usepackage[utf8]{inputenc} 
\usepackage[T1]{fontenc}
\usepackage{url}              
\usepackage{cite}             

\usepackage[cmex10]{amsmath}  
\usepackage{mleftright}       
\mleftright                   
\usepackage{amsthm}
\usepackage{graphicx}         
\usepackage{booktabs}         

\usepackage{cite}
\usepackage{amsmath}
\usepackage{amssymb}
\usepackage{algorithmic}
\usepackage{textcomp}
\usepackage{xcolor}
\def\BibTeX{{\rm B\kern-.05em{\sc i\kern-.025em b}\kern-.08em
    T\kern-.1667em\lower.7ex\hbox{E}\kern-.125emX}}
\usepackage{graphicx}
\usepackage{subfiles}
\usepackage[inline]{enumitem}

\usepackage{array} 
\usepackage{url}
\usepackage{epstopdf}
\usepackage{tikz}
\usetikzlibrary{automata,arrows,positioning,calc,
shapes,chains}
\usepackage{mathtools}
\usepackage{texdef2015}
\allowdisplaybreaks

\renewcommand{\vec}[1]{\begin{bmatrix} #1\end{bmatrix}}

\newcommand{\age}{\Delta}

\newlength{\swwidth}

\newcommand{\lam}{\lambda}

\begin{document}

\title{Age-Memory Trade-off in Read-Copy-Update} 


\author{\IEEEauthorblockN{Vishakha Ramani,
Jiachen Chen,
Roy D. Yates}
\IEEEauthorblockA{WINLAB, Rutgers University \\
Email: \{vishakha, jiachen, ryates\}@winlab.rutgers.edu
}}

\maketitle

\begin{abstract}
In the realm of shared memory systems, the challenge of reader-writer synchronization is closely coupled with the potential for readers to access outdated updates. Read-Copy-Update (RCU) is a synchronization primitive that allows for concurrent and non-blocking read access to fresh data. This is achieved through the creation of updated data copies, with each prior version retained until all associated read-locks are released. 
Given the principle that frequent updating keeps information fresh, the concern is whether we accumulate an infinite number of update copies, leading to excessively large memory usage.
This paper analyzes trade-offs between memory usage and update age within real-time status updating systems, focusing specifically on RCU. The analysis demonstrates that with finite read time and read request rate, the average number of updates within the system remains bounded. 
\end{abstract}

\section{Introduction}
Many real-time applications, commonly found on pervasive smartphones, are 
multi-threaded, prompting the critical question of identifying those 
mechanisms that facilitate timely  data transfers between processes within an operating system. For these Inter-Process Communication (IPC) mechanisms, shared memory has emerged as a notable contender 
\cite{ipcbench}.

However, the dichotomy inherent in memory systems, wherein writers and readers function in a state of mutual unawareness, creates an asynchronous operational paradigm. 
This asynchrony poses various challenges, with two key concerns: (i) the imperative to synchronize writers and readers to avoid race conditions, and (ii) the potential for readers to encounter stale updates due to temporal disparities between the writing and reading processes. 
These issues
warrant a closer examination regarding the timely retrieval of stored data items.

Furthermore, these issues are coupled, as using a lock-based synchronization primitive ensures that readers are blocked while the writer is writing a fresh update. Consequently, readers read the freshest data item in memory, but at the cost of increased read latency and an increased age on the reader's client side. Conversely, lock-less primitives such as Read-Copy-Update (RCU) \cite{mckenney1998read} provide non-blocking access to data but may result in potentially stale data being read, leading to potentially incorrect computations on the reader's client side. 

These issues were examined in \cite{ramani-cyINFOCOM, ramani-cy-AoI-INFOCOM}, particularly in the context of a timely packet forwarding application.
This prior work on RCU showed that while RCU reduces latency by enabling non-blocking readers and eliminating mutual exclusion among readers and writers, this improvement comes at the expense of increased memory usage. This becomes a particular concern when the application using such a primitive is running on a resource-constrained mobile device.

Consider a scenario of a Visual Simultaneous Localization and Mapping (SLAM) system \cite{Mur_Artal_2017},
which  constructs a map of an environment in real-time based on sensor data while simultaneously determining the location of a mobile device within that map. 
For seamless interaction with the real world, it is desirable to run SLAM systems on mobile phones.  In a typical SLAM workflow, incoming images are processed to track the device's location, and this location information is then incorporated into a global map, with ongoing optimization of the map structure. 
For timely accuracy,
SLAM systems must promptly process incoming camera streams, accessing the latest images in real-time. 
Although SLAM systems adopt a modular approach with concurrent modules handling 
specific tasks such as image processing, location tracking, map updating, and global map optimization,
there is a tight coupling between modules. All modules operate on the global map, implemented as a shared data structure, 
and engage in
computationally intensive operations, frequently accessing and updating this shared data structure \cite{sofia-HotMobile-2022, DantuEdgeSlam}.


RCU is particularly well-suited to  applications such as Visual SLAM because it enables a module to read the freshest copy of a data item. When a module performs a complex operation using a data item, it places a read-lock on that data to ensure it will be available and unchanged during the read operation. When the RCU writer wishes to update the data item, the write operation creates a fresher version/copy. As soon as the write is committed, this fresher copy is returned to subsequent read requests. However, each prior copy is retained in memory until all of its read-locks have been released.
Therefore, from a timeliness perspective, more frequent updating of data items in the memory provides the latest information to readers but this results in memory overhead by increasing the number of data copies created. 

While Visual SLAM serves as an illustrative example, the broader motivation is to explore the trade-off between memory usage and update age in real-time systems.
RCU, as a widely used synchronization primitive, is the focal point of our study in the following ways:
\begin{enumerate*}
    \item We investigate the memory footprint of concurrent updates in RCU and provide an 
    upper bound on the average number of active\footnote{An update is active if there is at least one reader reading that update.} updates in the system, and 
    \item we analytically explore a trade-off between memory footprint and the age of updates, particularly in case of unbounded number of concurrent updates. 
\end{enumerate*}

\subsection{Read-Copy-Update (RCU) Overview}

Replacing expensive conventional locking techniques, Read-Copy-Update (RCU) is a synchronization primitive that allows concurrent forward progress for both writers and readers \cite{MckenneyRCU2001}.  The operation of RCU has two key stages \cite{kernelRCU, guniguntala}:
\begin{enumerate*}
    \item \textit{Publishing a Newer Version:}
    To write a fresher version of a data item, the writer initiates the process by duplicating the RCU-protected data and subsequently modifying this duplicate with the fresher content. This modification occurs atomically, effectively replacing the old reference with a pointer to the new version. 
    \item \textit{Memory Reclamation and Deferred Deletion:} The publication of the modified data item marks the start of a  ``grace period'' that terminates when all existing RCU read-side critical sections  to have completed~\cite{Desnoyers}.
    Therefore, the end of grace period ensures that it is safe to reclaim the memory and delete the stale copy.
    
\end{enumerate*}

Notably, the  RCU publishing process runs concurrently with ongoing read operations, allowing the reads to persistently access the old version of an update using the original reference. However, new read requests, initiated after the publication, retrieve the most recent version of the data. Thus, for each data item, RCU maintains multiple time-stamped versions of each data item - a current version as well as a random number of prior ``stale'' versions in their respective grace periods.
 


\section{System Model and Main Results}
\label{sec:RCU-analysis}
\begin{figure}
    \centering
    \includegraphics[width=\linewidth]{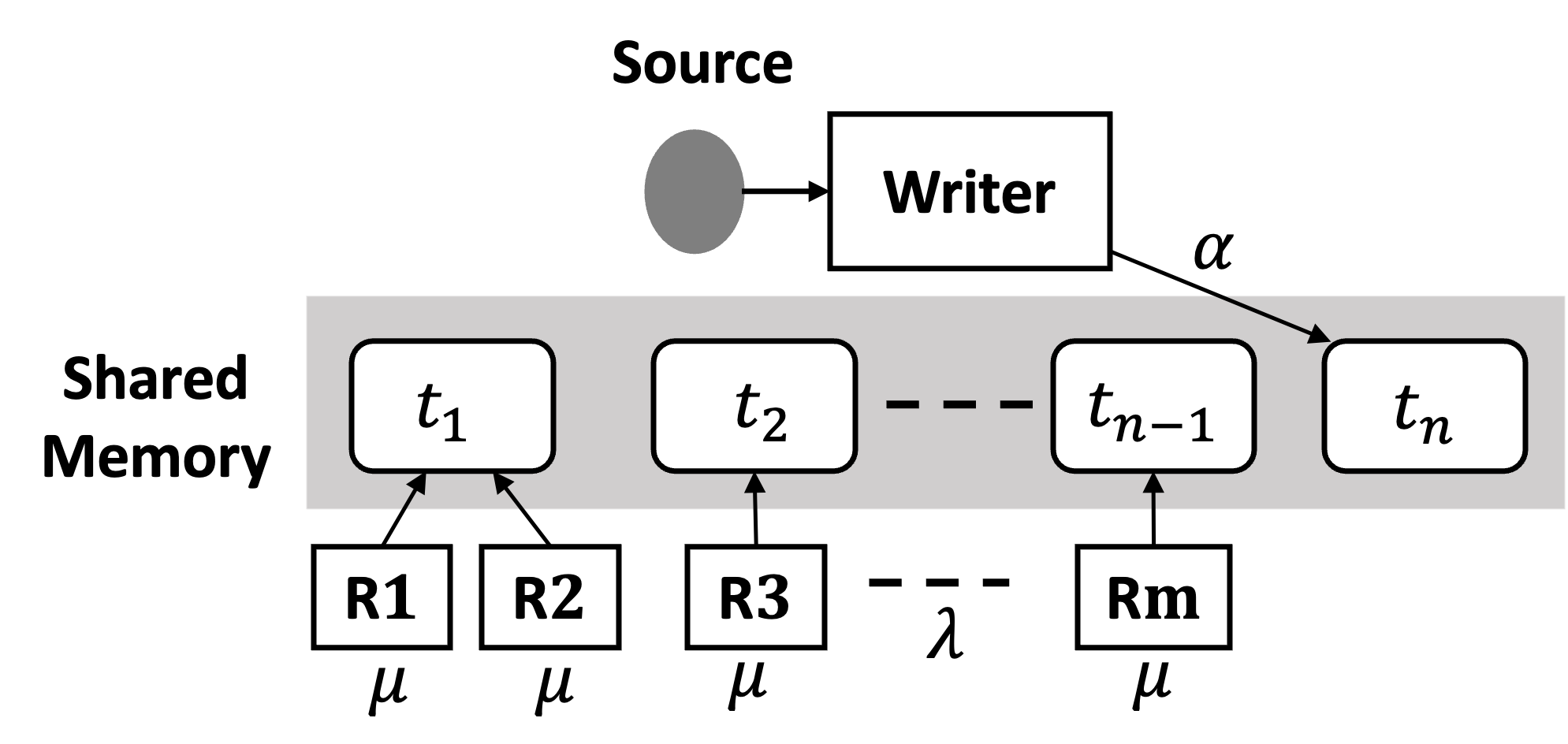}
    \caption{{\em Memoryless RCU model:} On behalf of an external source, a writer updates the shared memory at rate $\alpha$ with timestamped updates, denoted by timestamps $t_1, t_2, \ldots$.  Read requests $R_1, R_2, \ldots, R_m$  access the version of the source update with the freshest timestamp. These read requests are generated at rate $\lambda$ and have a mean read time of $1/\mu$.}
    \label{fig:agememtf-sysmod}
\end{figure}
In this work, we focus on a  class of systems (see Fig.~\ref{fig:agememtf-sysmod}) in which a source generates time-stamped {\em updates},
which are stored in shared memory. The writer queries this source for fresh measurements to update the memory, creating a new copy therein. Concurrently, a reader serves clients' requests for these measurements by accessing the memory. Multiple `old' readers may concurrently access distinct versions of data copies, depending upon the time of their request. It is noteworthy that the most recent read request will consistently retrieve the latest update from the memory.

In practical systems, a "preparation time" for update generation may exist in response to a writer's query. However, to emphasize the shared memory impact, we assume negligible preparation times throughout this work. Thus, the writer consistently receives fresh (zero age) updates from the source. Furthermore, although multiple sources may exist in the system, our attention centers on the shared data structure tracking the status of a single process of interest.

To analyze RCU, we assume the writer starts writing a fresh update as soon as it finishes its previous write, without regard for the number of update copies in the grace period.
With respect to memory consumption (i.e. the number of copies created), this is a worst-case analysis in that the writer is pushing to create as many copies as possible.
In practice, the number of update versions is limited by physical memory; however, we ignore this constraint here. Instead, we employ a model that limits the creation of copies by constraining how fast the writer can write an update to memory. In this regard, we will sometimes call such a writing process as {\em unconstrained write process}. Specifically, we examine a system in which write operations to unlocked memory have  independent exponential $(\alpha)$ service times. Since the writer receives a fresh update from the source immediately after publishing an update, there is a rate $\alpha$ Poisson point process of new updates being generated and written to memory.

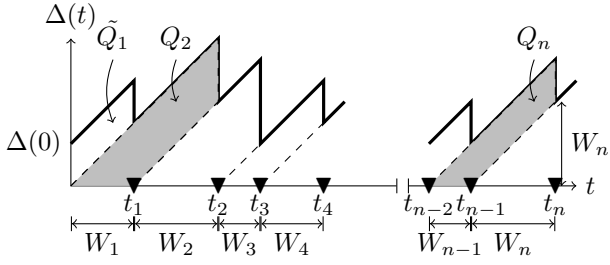
\begin{figure}[t]
\centering
\begin{tikzpicture}[scale=0.28] 
\draw [<-|] (0,7) node [above] {$\age(t)$} -- (0,0) -- (15.5,0);
\draw [|->] (16,0) -- (24,0) node [right] {$t$};
\draw [very thick] (0,2) -- (3,5) -- (3,3)  -- (7,7) 
-- (7,4)  -- (9,6) -- (9,2) -- (12,5) -- (12,3) -- (13,4);
\draw [fill=lightgray, ultra thin, dashed] (0,0) to (7,7) to (7,4) to (3,0);
\draw [ultra thin, dashed] (0,0) to (7,7) to (7,4) to (3,0);
\draw[<-] (2,3) to [out=110,in=250] (2,6) node [above] {$\tilde{Q_1}$};
\draw[<-] (5,4) to [out=110,in=250] (5,6) node [above] {$Q_2$};
\draw 
(0,2) node [left] {$\age(0)$}
(3,0) node {$\blacktriangledown$}
(3,0) node [below] {$t_1$} 
(7,0) node {$\blacktriangledown$}
(7,0) node [below] {$t_2$}
(9,0) node {$\blacktriangledown$}
(9,0) node [below] {$t_3$}
(12,0) node {$\blacktriangledown$}
(12,0) node [below] {$t_4$};
\draw  [|<->|] (0,-1.8) to node [below] {$W_1$} (3,-1.8);
\draw  [|<->|] (3,-1.8) to node [below] {$W_2$} (7,-1.8);
\draw  [|<->|] (7,-1.8) to node [below] {$W_3$} (9,-1.8);
\draw  [|<->|] (9,-1.8) to node [below] {$W_4$} (12,-1.8);
\draw [thin, dashed] (7,0) to (12,5) to (12,3) to (9,0);
\draw [ultra thin, dashed] (17,0) to (23,6) to (23,4) to (19,0);
\draw [very thick] (17,2) -- (19,4) -- (19,2) -- (23,6) -- (23,4) -- (24,5); 
\draw [fill=lightgray, ultra thin, dashed] (17,0) to (23,6) to (23,4) to (19,0);
\draw
(17,0) node {$\blacktriangledown$}
(17,0) node [below] {$t_{n-2}$}
(19,0) node {$\blacktriangledown$}
(19.5,0) node [below] {$t_{n-1}$}
(23,0) node  {$\blacktriangledown$}
(23,0) node [below] {$t_{n}$};
\draw[<-] (22,4) to [out=110,in=250] (22,6) node [above] {$Q_n$};
\draw  [|<->|] (17,-1.8) -- node [below] {$W_{n-1}$} (19,-1.8);
\draw  [|<->|] (19,-1.8) -- node [below] {$W_n$} (23,-1.8);
\draw [<->] (23.3,4) -- node [right] {$W_n$} (23.3,0); 
\end{tikzpicture}
\caption{Example evolution of age at shared memory
in the unconstrained write model. 
Updates are published 
in
memory at times marked $\blacktriangledown$.}
\label{fig:age}
\end{figure}

Fig.~\ref{fig:age} shows a sample age evolution process at shared memory as a function of time $t$. Without loss of generality, assume an update $0$ with initial age $\age(0)$ is in memory at time $t = 0$. 
Following the publication of an arbitrary update $n-1$ at time 
$t_{n-1}$, the writer queries the source for a fresh measurement. 
In response, the source
generates an 
update $n$ with time-stamp $t_{n-1}$. The writer receives this update instantly, 
begins writing to the shared memory, and subsequently publishes the new update at time $t_n$.

The age $\age(t)$ at the shared memory increases linearly in time in the absence of any new update and is reset to a smaller value when an update is published. Thus, at time $t_n$, $\age(t)$ is reset to $W_n = t_n - t_{n-1}$. This phenomenon continues for all subsequent updates 
and therefore, the age process $\age(t)$ exhibits a sawtooth waveform shown in Fig.~\ref{fig:age}. 
The time-average age is the area under graph in Fig.~\ref{fig:age} normalized by time interval of observation. A stationary ergodic age process $\age(t)$ has average age (often referred to as AoI)  \cite{yates2018ToIT}:
\begin{equation}
    \E{\age} = \lim_{T\to\infty}\frac{1}{T}\int_0^T\age(t)dt.
    \label{eq:agedefn}
\end{equation}

We further assume that the read requests form a rate $\lambda$ Poisson process, and each request's service/read time is an independent exponential $(\mu)$ random variable. 
Furthermore, we assume that memory is reclaimed when the last reader, holding the reference to a particular update, completes its service time.
We refer to this as the \textit{memoryless RCU} model since both write initiations and read requests  are memoryless Poisson point processes. 

Since read requests arrive as a Poisson process and the reads have exponential holding/service times independent of the number of concurrent read requests of an update, the birth-death process of read locks is an M/M/$\infty$ queue.  
However, from the perspective of the birth-death process of update copies in memory, the RCU system is complicated because update $n$ is tagged by a number of read requests that depends on the write time of update $n+1$. This implies that
the service/active time of an update depends upon the inter-arrival time of the next update; this is not an M/M/$\infty$ queue.


\subsection{Main Results}
Let $N(t), t \geq 0$ denote  the stochastic process of the number of active updates at time $t$. 
When each update has a fixed size in memory,  $N(t)$ is proportional to the memory footprint of the RCU updating process. 
Theorem \ref{thm:en-ub1} desribes the memory footprint $\E{N}$ and the average age $\E{\age}$ of an update in memory in terms of the system parameters $\lambda$, $\mu$ and $\alpha$.
\begin{theorem}\label{thm:RCU}
For the memoryless RCU model in which updates are written as a rate $\alpha$ Poisson process, and read requests arrive as a rate $\lambda$ Poisson process with independent exponential $(\mu)$ service times: 
\begin{enumerate}
\item[(a)] The memory footprint $\E{N}$ satisfies
\begin{align}
    \E{N} &=
     1 + \sum_{k = 1}^{\infty}
    \sum_{j=0}^\infty \frac{b_k^j e^{-b_k}}{j!}\parfrac{j}{\alpha/\mu + j}.\eqnlabel{EN2}
\end{align}
where $ b_k = \lambda q^k/\mu$ with $q =\alpha/(\alpha+\mu)$.
\item[(b)] \label{thm:en-ub1} 
\begin{align}
    \E{N} 
    &\leq 1 +  \sum_{k =1}^\infty \frac{q^k}{q^k+\alpha/\lambda}\eqnlabel{en-ub1}\\
&\le 1 + {\lambda}/{\mu}. \eqnlabel{en-ub2}
\end{align}

\item[(c)] \label{thm:avg-age} 
The average age of the current update in memory is
\begin{equation}
    \E{\age} = {2}/{\alpha}. \label{eq:avg-age}
\end{equation}
\end{enumerate}
\end{theorem}

\section{Proof of Theorem \ref{thm:en-ub1}}
\label{sec:proof1}
\subsection{Proof of Theorem~\ref{thm:RCU}(a)}
\begin{figure}[t]
    \centering
    \includegraphics[width=\linewidth]
    {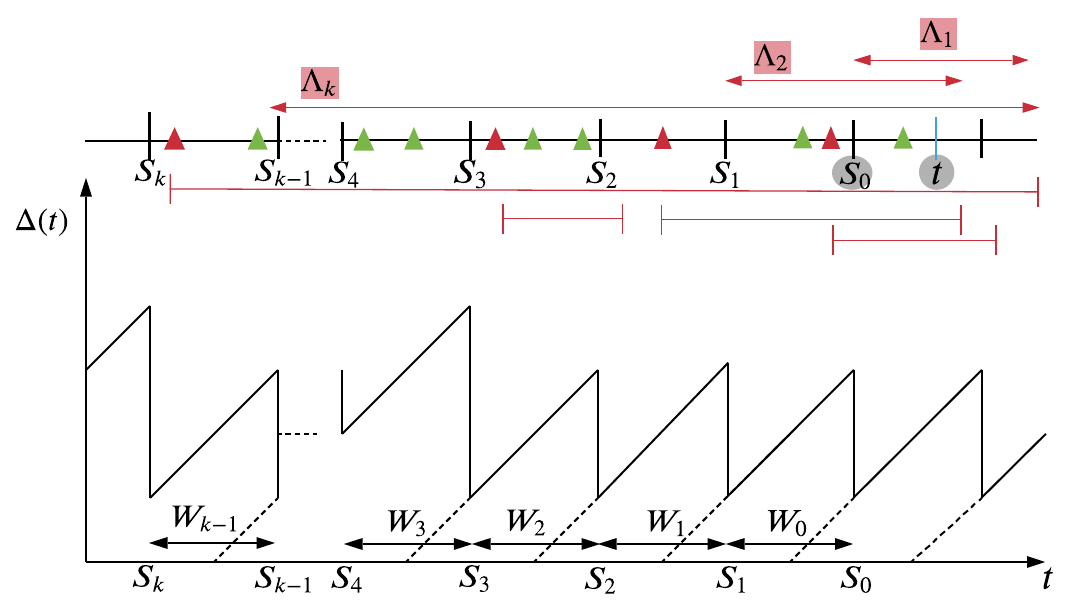}
    \caption{An example of the RCU read/write process (upper timeline) and the sample age evolution (shown only for illustration purpose) of update in memory (lower timeline). In the upper timeline: 
    green triangles mark arrivals of read requests that finish before the next update is published; red triangles mark those reads that establish a grace period by holding a read lock after the next update is published;
    the red intervals beneath the upper timeline show the service times of such readers; the red arrows above the upper timeline (with labels $\Lambda_k$, $\Lambda_2$ and $\Lambda_1$) identify the grace periods of updates $k$, $2$, and $1$ that are active at time $t$. 
    }
    \label{fig:uncons-cartoon}
\end{figure}

Consider an example of unconstrained write process as shown in Fig.~\ref{fig:uncons-cartoon} along with the corresponding age evolution. We inspect the system at an arbitrary time $t$. Relative to time $t$, we look backward in time and define update $0$ to be the most recently published update. We refer to update $0$ as the current update. We also set our clock  such that update $0$ is published at time $S_0=0$.  Further, we use index $k\ge0$ to denote the update published $k$ writes prior to update $0$. We denote publication time of update $k$ by $S_{k}$ and thus the $S_{k}$ are indexed backward in time, i.e., $\ldots,S_{n+1} < S_{n} < S_{n-1} < \ldots, S_{1} < S_0 = 0$. 
Following this notation, the writing time of  update $n$ is $W_n = S_{n-1} - S_{n}$. 
Recall that $W_n$ are i.i.d. exponential $(\alpha)$ random variables. 
By the memoryless property of the 
exponential random variable,
$Z = t - S_{0}$
,  the time elapsed since the last published update,  is also  exponential $(\alpha)$.  

When the writer publishes update $k-1$ at time $S_{k-1}$, the grace period for update $k$ starts, and the writer starts writing update $k-2$. At this time, $S_{k-1}$, there is a random number of residual reader locks on update $k$ and the grace period for update $k$ terminates when all these residual readers release their respective locks. 

%

For each past update $k>0$, there is some probability that it remains in a grace period at time $t$.
Given $W_{k-1} = w$, update $k$ is the current update in the interval $(S_{k},S_{k-1})=(S_{k-1}-w,S_{k-1})$. In this length $w$ interval, the number of read requests $M_k$ is Poisson with $\E{M_k}=\lambda w$. Moreover, given $M_k=m$, the arrival times of the read requests are statistically identical to the set $\set{S_{k-1}+U_1,\ldots,S_{k-1}+U_m}$, where $\seq{U}{1}{m}$ is a set of i.i.d.~uniform $(-w,0)$ random variables  \cite{Ross}. These read requests will have i.i.d.~exponential $(\mu)$ service times $X_1,X_2,\ldots,X_m$. 
The $i$th such read request releases its  read-lock at time $S_{k-1}+ Y_i$ where $Y_i =U_i+ X_i$. 
Therefore, given $M_k = m$ and $W_{k-1} = w$ the last read-lock on update $k$ is released at time  
\begin{equation}
    \Lambda_k = S_{k-1}+\max_{1\le i\le m} Y_i.
\end{equation}
We define $L_{k-1}$ as the time elapsed since $S_{k-1}$ up to time $t$. 
Then,
\begin{equation}
    L_{k-1} = \sum_{j = 0}^{k-2} W_{j} + Z.
    \eqnlabel{fLk-1}
\end{equation}
The number of active updates at time $t$, $N$,  is equal to the number of updates still in their respective grace periods at time $t$ plus the current published update $0$. To find $\E{N}$, we define $E_k$ as the event that the grace period of update $k$ has ended by time $t$. 
%
The conditional probability of event $E_k$ that update $k$ has finished its service by time $t$ is
\begin{align}
    &\hspace{-2.8em}\prob{E_k \mid W_{k-1} = w,  M_k = m} \nn
    &= \prob{\Lambda_k \leq  S_{k-1}+ L_{k-1} \mid W_{k-1} = w, M_{k} = m}\nonumber \\
    & = \prob{\max_{1\le i\le m}Y_i \leq L_{k-1} \mid W_{k-1} = w}\nn
    & = \paren{\prob{Y_i \leq L_{k-1} \mid W_{k-1} = w}}^m, 
\end{align}
where we have used the fact that the $Y_i=U_i+X_i$ remain i.i.d.~under the condition $W_{k-1}=w$.  We observe that $L_{k-1}$ and $X_i$ are independent of $W_{k-1}$. For fixed $w$, we also observe that $U_i$ depends on $W_{k-1}$ only to the extent that the event $W_{k-1}=w$ specifies that $U_i$ is a uniform $(-w,0)$ random variable. Hence,  defining $X$ to be exponential $(\mu)$ and $U$ to be uniform $(-w,0)$,
\begin{align}
    \prob{E_k \mid W_{k-1} = w,  M_k = m} 
    &=\paren{\prob{U+X \leq L_{k-1}}}^m.\eqnlabel{UXL}
    \end{align}

\begin{lemma}
\begin{equation}
    \prob{U+X \leq L_{k-1}} = 1 - a_w q^k = \epsilon(k,w) \eqnlabel{y-less-l-short}
\end{equation}
where $q =\alpha/(\alpha+\mu)$ and
$a_w = (1- e^{-\mu w})/\mu w$.
\label{lemma:y-less-lk}
\end{lemma}
The proof, an elementary probability exercise,  appears in the Appendix.
It then follows from  \eqnref{UXL} and  \eqnref{y-less-l-short} that 
\begin{align}
    &\hspace{-1.2em}\prob{E_k \mid W_{k - 1} = w} \nn
    &= \sum_{m = 0}^{\infty} \prob{E_k\mid W_{k - 1} = w,M_k=m} \pmf{M_{k} \mid W_{k-1}}{m \mid w}, \nonumber \\
    &= \sum_{m = 0}^\infty \frac{1}{m!}\epsilon(k, w)^m (\lambda w)^m e^{-\lambda w}
    = \exp\left(-b_k (1 - e^{-\mu w})\right),
    \eqnlabel{conditional-p-ek}
    \end{align}
where $b_k = \lambda q^k/\mu.$
Since, $W_{k-1}$ is exponential $(\alpha)$, it follows from \eqnref{conditional-p-ek} that
\begin{align}
     \prob{E_k} &= \int_{0}^{\infty} P(E_k \mid W_{k-1} = w) f_{W_{k-1}}(w) \,dw \nn
    &= \alpha \int_{0}^\infty e^{-b_k (1-e^{-\mu w})} e^{-\alpha w} \,dw.  \label{eq:p_ek-actual}
\end{align}
With the substitution $y=e^{-\mu w}$, we obtain \begin{align}
    \prob{E_k}
    &= \frac{\alpha e^{-b_k}}{\mu} \int_{0}^{1} y^{\alpha/\mu - 1}e^{b_k y} \,dy. 
\end{align}
A Taylor series expansion of $e^{b_ky}$ yields
\begin{align}
    \prob{E_k}
    &=\frac{\alpha e^{-b_k}}{\mu} \int_0^1 y^{\alpha/\mu -1}\sum_{j=0}^\infty \frac{(b_ky)^j}{j!}\,dy\nn
   &= \frac{\alpha e^{-b_k}}{\mu}
   \sum_{j=0}^\infty \frac{b_k^j}{j!}\int_0^1 y^{\alpha/\mu+j-1}\,dy, \nn
   &= \frac{\alpha e^{-b_k}}{\mu} \sum_{j=0}^\infty \frac{b_k^j}{j!(\alpha/\mu +j)}.
   \end{align}
It follows that 
\begin{equation}
    \prob{E_k^c} =  1-\prob{E_k}
    =\sum_{j=0}^\infty \frac{b_k^j e^{-b_k}}{j!}\parfrac{j}{\alpha/\mu + j}.\eqnlabel{PEkc}
\end{equation}
    
Now let $I_{k}$ be the indicator random variable for the event $E_k^c$ that update $k$ is active at time $t$. 
Therefore, the number of active updates is 
$N = 1 + \sum_{k = 1}^{\infty} I_{k}$.  
Hence,
\begin{align}
    \E{N} &= 1 + \E{\sum_{k = 1}^{\infty} I_{k}}
    = 1 + \sum_{k = 1}^{\infty} \mathbb{P}(E_k^c). \eqnlabel{EN1}
\end{align}
Theorem~\ref{thm:en-ub1}(a) follows from \eqnref{PEkc} and \eqnref{EN1}.

\subsection{Proof of Theorem~\ref{thm:RCU}(b)} 
To verify Theorem~\ref{thm:en-ub1}(b), let $J_k$ denote a Poisson $(b_k)$
random variable. We observe that \eqnref{EN2} can be written as
\begin{align}
    \E{N} &=
     1 + \sum_{k = 1}^{\infty}
    \E{\frac{J_k}{\alpha/\mu +J_k}}.\eqnlabel{EN3}
\end{align}
Since $x/(\alpha/\mu+x)$ is a concave function, it follows from 
Jensen's inequality that
\begin{align}
    \E{N} &\le
     1 + \sum_{k = 1}^{\infty}
    \frac{\E{J_k}}{\alpha/\mu +\E{J_k}}
    =1 + \sum_{k = 1}^{\infty}
    \frac{b_k}{\alpha/\mu +b_k}, \nn
    &=1 + \sum_{k = 1}^{\infty}
    \frac{q^k}{\alpha/\lambda +q^k}.\eqnlabel{ENUB1}
\end{align}
Since $q^k\ge 0$, it follows from \eqnref{ENUB1} that 
\begin{align}
    \E{N}\le 1+\sum_{k=1}^\infty \frac{\lambda}{\alpha}q^k
    =1+\frac{\lambda q}{\alpha(1-q)}
    =1+\frac{\lambda}{\mu}.
    \eqnlabel{ENUB2}
\end{align}

\subsection{Proof of Theorem~\ref{thm:RCU}(c)}
Fig.~\ref{fig:age} represents a sample age evolution in the unconstrained write model with $W_n$ denoting the exponential $(\alpha)$ write time of the $n$th update. We represent the area under sawtooth waveform as the concatenation of the polygon areas $\tilde{Q_{1}}, Q_{2}, \ldots, Q_{n}, \ldots$. The average age is
\begin{align}\eqnlabel{EQoverEW}
    \age = \frac{\E{Q_n}}{\E{W_n}}
\end{align}
where
\begin{align}
    Q_{n} &= \frac{(W_{n-1} + W_{n})^{2}}{2} - \frac{W^{2}_{n}}{2}
 = \frac{W_{n-1}^2 + W_{n-1}W_{n}}{2}.
\end{align}
Since $\E{W_n}=1/\alpha$ and $\E{W_n^2}=2/\alpha^2$, 
\begin{align}
    \E{Q_{n}} &= \frac{1}{2}\E{W_{n-1}^2} + \E{W_{n-1}}\E{W_n}
    = \frac{2}{\alpha^2}\eqnlabel{EQn}
\end{align}
 and the claim follows from \eqnref{EQoverEW} and \eqnref{EQn}.

\section{Numerical Evaluation and Discussion}
\begin{figure}[t]
    \centering
    \begin{tabular}{c}
    \includegraphics{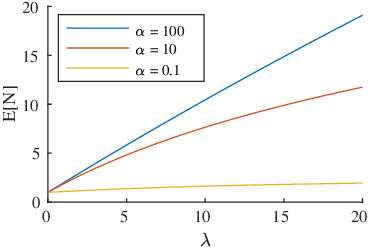} \\
    \textbf{(a)} \\
    \includegraphics{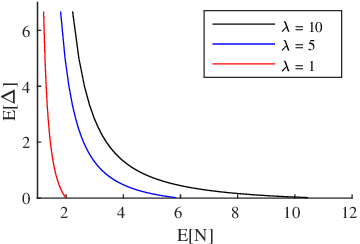}
    \\
    \textbf{(b)}
    \end{tabular}
    \caption{(a) Memory footprint in RCU as a function of read arrival rate $\lambda$. (b) Trade-off between the average age $\age$ and $\E{N}$ as a function of writing rate $\alpha$. In both $(a)$ and $(b)$, the read service rate is $\mu=1$.}
    \label{fig:mem-trade-off}
\end{figure}

From Theorem~\ref{thm:avg-age}, we see that the average age $\E{\age}$ of the current update in the memory  is monotonically decreasing with the writing rate $\alpha$. Fig.~\ref{fig:mem-trade-off}(b) plots 
age-memory trade-off  over $\alpha \in (0, \infty)$, showing that minimal average age at the readers is achieved when 
updates are written as fast as possible, but this is at the expense of an increased memory footprint. However, for a fixed write rate $\alpha$, the memory footprint is an increasing function of the read request rate $\lambda$ as shown in Fig.~\ref{fig:mem-trade-off}(a).
Both the analysis and numerical evaluation highlight the trade-off between age and memory observed in the RCU mechanism.

\begin{figure}[t]
    \centering
    \begin{tabular}{c}
         \includegraphics{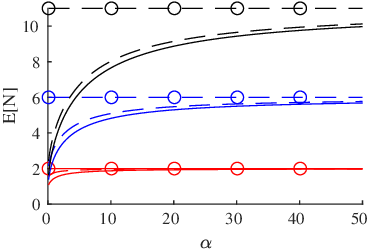}
         \\
         \textbf{(a)} \\
         \includegraphics{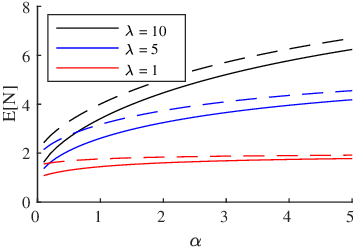} 
         \\
         \textbf{(b)}     
    \end{tabular}
    \caption{(a) The expected number of active updates are written at rate $\alpha$. The black, blue and red curves are when $\lam/\mu = 10$, $\lam/\mu = 5$, and  $\lam/\mu = 1$ respectively; the read service rate is $\mu=1$. (b) Zoomed in version of (a)).}
    \label{fig:en-ub}
\end{figure}

Fig.~\ref{fig:en-ub} plots $\E{N}$, and the upper bounds 
\eqnref{en-ub1} and \eqnref{en-ub2} as a function of $\alpha$ for $\mu = 1$ and various $\lambda$.
We observe that the upper bound \eqnref{en-ub1} is tight for all $\alpha$. Further, notice that as $\alpha \to \infty$, the expected number of updates for different values of $\lambda$ approach the upper bound in \eqnref{en-ub2}, albeit at different rates.

We now give some intuition for  the upper bound to $\E{N}$ in Theorem~\ref{thm:en-ub1}(b).
For $\alpha\gg\lambda$, each update is tagged with zero or one reads. As $\alpha\to\infty$, an untagged update expires in expected time $1/\alpha\to 0$, as it is replaced by the next update.  On the other hand, a tagged update enters a grace period with duration corresponding to the exponential $(\mu)$ service time required by its read. Hence tagged updates have a one-to-one correspondence with the reads in the system. The number of tagged updates is described by the M/M/$\infty$ queue process with arrival rate $\lambda$ and service rate $\mu$ that characterizes the number of reads in the system. 
Therefore, in the limiting case of $\alpha\to\infty$, the number $N'$ of tagged updates in the system follows a Poisson distribution
\begin{align}
    \prob{N'=n} = \frac{(\lambda/ \mu)^n e^{-\lambda / \mu}}{n!}, \quad n \geq 0.
\end{align} 
Furthermore, there is always one untagged update that is perpetually being replaced. 
Hence, the number of updates in the system is $N=1+N'$ and the average number of updates holding a read lock is 
    $\E{N}=\E{1+N'} = 1 + {\lambda}/{\mu}$.

\section{Conclusion}
In this work, we explored the trade-off between memory footprint and update age in the context of RCU, particularly relevant for applications with sizable updates operating within the constraints of memory-constrained mobile devices. 
The central question revolves around whether frequent updating 
can induce excessive memory consumption. 
Theorem~\ref{thm:RCU} provides a reassuring finding — given finite average service/read time $1/\mu$  and read request rate $\lambda$, the average number of updates in the system is finite.

One way the memory usage can be regulated is by limiting the update rate $\alpha$ of the writer, albeit at the expense of high AoI. 
Alternatively, the memory footprint can be reduced 
by controlling the read request rate $\lambda$. 
Consider a setting in which an application (such as SLAM) processes an update in a number of modules that can be executed either concurrently or  sequentially.  
In this setting, sequential operation
can effectively reduce the read request rate $\lambda$ through various methods.
For example, the modules may generate individual read requests,  but the sequential execution of the modules will slow the overall update processing rate. In an alternate approach that utilizes local copies, the application makes a single read to store the current data item  in a local copy. This local copy is then utilized by  the subsequent module executions. Although a caveat can be that for large objects in the memory, the reads take longer. If a process maintains a local copy, subsequent steps circumvent this read latency, but the data used will be stale if the main memory has been updated. 

While this work analyzes the tradeoff between the memory footprint and the age of updates in the memory, timeliness analysis of these proposals  for structuring the execution of the updating application is likely to be application-specific and remains as future work.

\bibliographystyle{IEEEtran}
\bibliography{refs, AOI-2020-03}

\appendix
\label{sec:lemma1-proof}
\begin{proof}
Since the  $W_{j}$ and $Z$ are i.i.d. exponential $(\alpha)$ random variables, \eqnref{fLk-1} implies  $L_{k-1}$ has a Gamma distribution  with PDF
\begin{equation}
    f_{L_{k-1}}(l) = \frac{\alpha}{\Gamma(k)}(\alpha l)^{k-1} e^{-\alpha l} \mathbf{1}_{\{l \geq 0 \}}.
\end{equation}
Since $Y = U + X$, where $U \sim \text{Uniform}(-w, 0)$ and $X \sim \exp{(\mu)}$, the PDF of $Y$ is 
\begin{IEEEeqnarray}{rCl}
    f_{Y} (y) &=& \int_{-\infty}^{\infty} f_X(x) f_{U}(y - x) \,dx \nonumber \\ 
    &=& \int_{-\infty}^{\infty} \mu e^{-\mu x} \mathbf{1}(x \geq 0) \frac{1}{w} \mathbf{1}(-w \leq y - x \leq 0) \, dx.\IEEEeqnarraynumspace
\eqnlabel{conditional-y}
\end{IEEEeqnarray}
Resolving the indicator functions in \eqnref{conditional-y} yields
\begin{align}
    f_{Y}(y) &= 
    \begin{cases} 
    \frac{1}{w}(1 - e^{-\mu(w+y)})
    ,\qquad -w \leq y \leq 0,\\
    \frac{1}{w}(e^{-\mu y} - e^{-\mu(w+y)}), \qquad y \geq 0.
    \end{cases}
\end{align}
This implies
\begin{align}
    \prob{Y \leq L_k} &= \int_{l = 0}^{\infty} f_{L_k}(l) \int_{-w}^{l}f_{Y}(y) \,dy\,dl 
    = I_1+I_2
\label{eq:y-less-l}
\end{align}
such that 
\begin{align}
    I_1 &= \int_{0}^{\infty} f_{L_k} (l) \int_{-w}^{0} f_{Y}(y) \,dy\,dl \nonumber \\
    &= \int_{0}^{\infty} f_{L_k} (l)\left[\int_{-w}^{0}  \frac{1}{w} \,dy  - \frac{1}{w} \int_{-w}^{0} (1- e^{-\mu(w+y)}) \,dy \right] \,dl  \nonumber\\
    &= 1 + 
(e^{-\mu w} - 1)/(\mu w),
\eqnlabel{integral-1}\\
    I_2 &= \int_{0}^{\infty} f_{L_k} (l) \int_{0}^{l} f_{Y}(y) \,dy\,dl, \nonumber \\
    &= - \underbrace{\frac{1}{\mu w} \int_{l = 0}^{\infty} f_{L_k} (l) (e^{-\mu l} - 1) \,dl}_{I_3} \nn
    &\qquad + \underbrace{\frac{1}{\mu w} \int_{l = 0}^{\infty} f_{L_k} (l) (e^{-\mu(w + l)} - e^{-\mu w}) \,dl.}_{I_4} 
\eqnlabel{integral-2}    
\end{align}
Solving integrals $I_3$ and $I_4$:
\begin{align}
    I_3 &= \frac{1}{\mu w} \left[\int_{l = 0}^{\infty} f_{L_k}(l) e^{-\mu l}dl  - 
    \int_{l = 0}^{\infty} f_{L_k}(l) dl\right] \nonumber \\
    & = \frac{1}{\mu w} \left[\int_{l = 0}^{\infty} \frac{\alpha}{\Gamma(k)} (\alpha y)^{k-1} e^{-(\alpha + \mu) l} dl - 
    1\right]
    \nn
    &= \frac{1}{\mu w} \left[(\frac{\alpha}{\alpha + \mu})^k - 1 \right], \label{eq:integral-3}\\
    I_4 &= \frac{1}{\mu w}\left[\int_{l = 0}^{\infty} f_{L_k}(l) e^{-\mu(w + l)}dl - \int_{l = 0}^{\infty} f_{L_k}(l) e^{-\mu w} dl \right] \nonumber \\
    &= \frac{1}{\mu w}\left[e^{-\mu w}\int_{l = 0}^{\infty}\frac{\alpha}{\Gamma(k)} (\alpha y)^{k-1} e^{-(\alpha + \mu) l} dl - e^{-\mu w} \right] \nn
    &= \frac{e^{-\mu w}}{\mu w} \left[(\frac{\alpha}{\alpha + \mu})^k - 1 \right].
    \label{eq:integral-4}
\end{align}
Combining \eqnref{integral-1}, \eqnref{integral-2},
(\ref{eq:integral-3}), and (\ref{eq:integral-4}), we have,
\begin{equation}
    \prob{Y \leq L_k \mid W_{k-1} = w} = 1 - \frac{1 - e^{-\mu w}}{\mu w} \left(\frac{\alpha}{\alpha + \mu}\right)^k.
\end{equation}
Recalling $q=\alpha/(\alpha+\mu)$ and 
$a_w = (1- e^{-\mu w})/(\mu w)$, 
the lemma follows. 
\end{proof}

\end{document}